\documentclass[sigconf,nonacm]{acmart}
\renewcommand\footnotetextcopyrightpermission[1]{}

\usepackage{booktabs}
\usepackage{listings}
\usepackage{xspace}

\usepackage{xcolor}
\newcommand{\TOTREPOS}{356}
\newcommand{\TOTCONFIGFILES}{612}
\newcommand{\TOTFILES}{9{,}470}

\newcommand{\TOTCONTEXTREPOS}{4{,}463}
\newcommand{\CANDIDATEELEMS}{29{,}454}
\newcommand{\CANDIDATEFILES}{8{,}213}
\newcommand{\CANDIDATEREPOS}{4{,}420}
\newcommand{\STALEREPOS}{82}
\newcommand{\STALEREPOSPCT}{23.0}
\newcommand{\STALEREPOSCILO}{18.8}
\newcommand{\STALEREPOSCIHI}{27.2}
\newcommand{\VALIDELEMS}{18{,}048}
\newcommand{\STALEELEMS}{230}
\newcommand{\STALEPCT}{1.27}
\newcommand{\STILVALID}{17{,}818}
\newcommand{\STILVALIDPCT}{98.73}

\newcommand{\toolname}{DOCER\xspace}

\begin{document}

\sloppy

\title[Context Rot in AI-Assisted Software Development]{Context Rot in AI-Assisted Software Development: Repurposing Documentation Consistency for AI Configuration Artifacts}

\author{Christoph Treude}
\affiliation{\institution{Singapore Management University}\city{Singapore}\country{Singapore}}
\email{ctreude@smu.edu.sg}

\author{Sebastian Baltes}
\affiliation{\institution{Heidelberg University}\city{Heidelberg}\country{Germany}}
\email{sebastian.baltes@uni-heidelberg.de}

\begin{abstract}
Developers increasingly provide AI coding assistants with persistent context through configuration files such as \texttt{CLAUDE.md}, \texttt{AGENTS.md}, and \texttt{.cursorrules}. These files describe code elements, architecture, and development conventions, forming the context that guides AI tool behavior across sessions. As software evolves, this context can become stale, a phenomenon we call \emph{context rot}. While AI configuration artifacts are new, the underlying consistency problem connects to decades of software documentation research. Researchers have built tools to check consistency between documentation and code, spanning README files, code comments, API documentation, architecture descriptions, and installation instructions. We argue that this existing toolbox is an immediate starting point for detecting context rot, and we present a research roadmap mapping documentation consistency approaches to corresponding problems in this new setting. As preliminary evidence, applying an existing README/wiki consistency checker to a statistically representative sample of \TOTREPOS{} repositories identifies stale code element references in \STALEREPOSPCT{}\% of repositories, showing that traditional documentation consistency tools can already surface context rot.
\end{abstract}

\keywords{AI coding assistants, documentation consistency, configuration artifacts, context rot}

\maketitle

%%% ============================================================
\section{Introduction}
%%% ============================================================

AI coding assistants are now a routine part of software development workflows. Tools such as Claude Code, GitHub Copilot, Cursor, and Gemini CLI generate code, answer architectural questions, and work through unfamiliar codebases on behalf of developers. Central to their effectiveness is persistent, project-specific context, supplied by configuration files that describe the codebase to the model before any interaction begins.

These configuration files take many forms, and most tools support multiple formats~\cite{galster2026configuring}. Anthropic's Claude Code reads \nolinkurl{CLAUDE.md} files placed at the repository root and in subdirectories. OpenAI's Codex reads \nolinkurl{AGENTS.md} and \nolinkurl{AGENTS.override.md}. GitHub Copilot reads \nolinkurl{.github/copilot-instructions.md} and \nolinkurl{instructions/*.md} files, and additionally recognizes \nolinkurl{CLAUDE.md} and \nolinkurl{AGENTS.md}. Cursor reads \nolinkurl{AGENTS.md} and formerly \nolinkurl{.cursorrules}, which has since been deprecated in favor of \nolinkurl{AGENTS.md}. Google's Gemini CLI reads \nolinkurl{GEMINI.md}. Despite their different naming conventions, the function is the same. Each gives the model project-specific knowledge that persists across sessions.
% GROUNDING: galster2026configuring -- "Exploratory study of configuring agentic AI coding tools; tools expose multiple configuration mechanisms and file formats (context files, rules, skills, hooks)." (arXiv 2602.14690)

These files describe many aspects of a project: conventions, contribution guidelines, architecture, build commands, and testing practices~\cite{mohsenimofidi2025context}. A \texttt{CLAUDE.md} might state that authentication logic lives in \texttt{src/auth/middleware.py}, that the primary API client is \texttt{HttpClient}, or that the project uses \texttt{MAX\_RETRIES} as a global constant. Beyond specific code references, these files also describe architectural patterns, required tools, dependency versions, and workflow expectations. \citeauthor{lulla2026impact} show that the presence of \texttt{AGENTS.md} files is associated with lower agent runtime and reduced token consumption~\cite{lulla2026impact}, suggesting that up-to-date configuration contributes to more efficient AI assistance. When stale, such descriptions can degrade AI assistance without any visible error. The model references a function that no longer exists, generates code that imports a deleted module, or enforces conventions the team abandoned months ago.
% GROUNDING: mohsenimofidi2025context -- "Context-engineering practices for AI agents in open-source software; configuration files capture conventions, contribution guidelines, architecture, build commands, and testing." (MSR '26)
% GROUNDING: lulla2026impact -- "Presence of AGENTS.md associated with lower median agent runtime (-28.64%) and reduced output token consumption (-16.58%) at comparable task completion." (arXiv 2601.20404, JAWs '26)

% NOTE: ``context rot'' is also used for a distinct phenomenon (Hong et al./Chroma 2025: LLM output
% degrading as the input context window grows). The footnote below states the distinction for readers.
We call this \emph{context rot}\footnote{The term \emph{context rot} is currently used mainly for degradation within a model's \emph{input context window}~\cite{hong2025contextrot}. We extend it to AI-specific, repository-versioned artifacts that are sent to models as project context or configure how AI coding tools operate. Their usefulness can degrade as they become stale relative to the evolving codebase.}: the gradual divergence between what a configuration file says about a codebase, its tools, its architecture, or its conventions, and what actually holds. While AI configuration artifacts are new, the underlying problem is not. The software engineering community has studied documentation consistency for decades, building tools to detect precisely this kind of mismatch: between code and README files~\cite{tan2024detecting}, between method implementations and their comments~\cite{tan2007icomment,tan2012tcomment,panthaplackel2021deep}, between APIs and their documentation~\cite{zhou2017analyzing,uddin2015api}, between architectural descriptions and deployed systems~\cite{ali2018architecture,keim2023detecting}, and between installation instructions and build configurations~\cite{borovits2022findici,gao2025adapting}. These tools predate AI coding assistants, but the consistency problems are structurally similar: take a claim made in a text artifact about the state of software, check it against the actual software, and flag divergences.
% GROUNDING: tan2024detecting -- "DOCER detects outdated code element references in repository documentation (README/wiki) via regex extraction + presence checks across repository snapshots." (EMSE 29(1):5, 2024)
% GROUNDING: tan2007icomment,tan2012tcomment,panthaplackel2021deep -- "iComment, @tComment, and deep just-in-time detection flag mismatches between code and its comments (inline / Javadoc)."
% GROUNDING: zhou2017analyzing,uddin2015api -- "API documentation defects: directive defects between docs and code (Zhou et al.); how API documentation fails (Uddin & Robillard)."
% GROUNDING: ali2018architecture,keim2023detecting -- "Architecture consistency state of practice/requirements (Ali et al.); traceability-link recovery to detect architecture-documentation inconsistencies (Keim et al.)."
% GROUNDING: borovits2022findici,gao2025adapting -- "FindICI detects code/NL inconsistencies in Infrastructure-as-Code (Borovits et al.); adapting installation instructions in evolving ecosystems (Gao et al.)."

This paper makes three contributions. First, we introduce context rot as the divergence between AI configuration files and the codebases, tools, architectures, and workflows they describe. Second, we provide preliminary empirical evidence for the feasibility of detecting context rot by applying an existing README/wiki consistency checker to AI configuration files without modification, finding stale code element references in \STALEREPOSPCT{}\% of \TOTREPOS{} repositories, sampled to be representative of the eligible dataset at a 95\% confidence level with a 5\% margin of error. Third, we present a research roadmap mapping five classes of documentation consistency technique onto open problems in AI configuration artifacts.

Specifically, we instantiate one part of a broader research roadmap: referential rot, where a configuration file refers to code elements that no longer exist. We apply \toolname{}~\cite{tan2024detecting}, an existing consistency checker for README files and wikis, to AI configuration files without adapting its extraction rules or staleness definition. Our aim is not to advance \toolname{} itself, but to test whether a traditional documentation consistency tool can identify stale code references in AI configuration artifacts.
% GROUNDING: tan2024detecting -- "We reuse DOCER's published regular-expression extraction and two-snapshot staleness definition without modification (EMSE 29(1):5, 2024)."

The remainder of this paper is organized as follows. Section~\ref{sec:background} provides background on AI configuration artifacts and the documentation consistency literature. Section~\ref{sec:study} describes the preliminary study and results. Section~\ref{sec:agenda} presents the research roadmap. Section~\ref{sec:conclusion} concludes.

%%% ============================================================
\section{Background}
\label{sec:background}
%%% ============================================================

We situate our work in two bodies of prior art: the emerging practice of AI configuration artifacts, and decades of research on software documentation consistency.

\subsection{AI Configuration Artifacts}

AI coding tools offer multiple configuration mechanisms, including context files, skills, subagents, and hooks~\cite{galster2026configuring}. Context rot can arise in any of these artifact types. Our preliminary study focuses on \emph{context files} — versioned Markdown files that developers maintain to provide persistent, project-specific context to an AI coding assistant — as the most widely studied artifact type.
% GROUNDING: galster2026configuring -- "Catalogs the configuration mechanisms across agentic AI coding tools (context files, rules/skills, subagents, hooks)." (arXiv 2602.14690)
Unlike interactive prompts, these files live in the repository alongside source code and are read automatically at the start of each session. Their content ranges from high-level architecture descriptions to precise references to specific functions, files, and constants.

\citeauthor{galster2026dataset} collected a dataset of \TOTFILES{} context files from \TOTCONTEXTREPOS{} GitHub repositories, spanning all major AI coding tool formats~\cite{galster2026dataset}.
% GROUNDING: galster2026dataset,baltes_2026_20090356 -- "context_files.csv: 9,470 context files across 4,463 distinct repos, within the 4,738 repos that have >=1 config artifact. Archived on Zenodo (doi:10.5281/zenodo.19375880)." (arXiv 2605.08435; Zenodo record)
The dataset reveals that these files are substantive: they are not mere pointer files but contain detailed descriptions that AI tools consume directly. The most common types are \texttt{AGENTS.md} (42.7\% of files), \texttt{CLAUDE.md} (30.3\%), and \texttt{copilot-instructions.md} (13.7\%). The content and purpose of such files have been examined from several angles: \citeauthor{mohsenimofidi2025context}~\cite{mohsenimofidi2025context} study context engineering practices for AI agents in open-source software, \citeauthor{galster2026configuring}~\cite{galster2026configuring} conduct an exploratory study of configuration practices across tools, and \citeauthor{lulla2026impact}~\cite{lulla2026impact} show empirically that \texttt{AGENTS.md} files improve AI coding agent efficiency, making the accuracy of these files relevant.
% GROUNDING: mohsenimofidi2025context -- "Studies context engineering practices for AI agents in open-source software." (MSR '26)
% GROUNDING: galster2026configuring -- "Exploratory study of configuration practices across agentic AI coding tools." (arXiv 2602.14690)
% GROUNDING: lulla2026impact -- "Shows empirically that AGENTS.md improves AI coding agent efficiency (lower runtime and token use)." (arXiv 2601.20404)

\subsection{Documentation Consistency Research}

Software documentation consistency has been an active research area since at least the early 2000s. The central challenge is that documentation and code evolve at different rates and through different processes. Code changes are continuously exercised by compilers, automated tests, and continuous integration, so drift is caught quickly. Documentation changes are not, so documentation is updated manually, opportunistically, and often incompletely.

Researchers have addressed consistency across multiple documentation artifact types, each targeting a distinct class of divergence. \toolname{}~\cite{tan2024detecting} targets README files and GitHub wikis, addressing stale code element references by extracting identifiers using regular expressions and checking their presence in the source code at successive repository snapshots. A reference is stale if the element was present when the documentation was first committed but has since been removed. Code comment consistency checkers address inconsistencies between code and comments, detecting when a method's implementation diverges from its Javadoc or inline comment~\cite{tan2007icomment,tan2012tcomment,panthaplackel2021deep}. API documentation checkers address API contract violations, comparing parameter names, types, and descriptions in published documentation against actual method signatures~\cite{zhou2017analyzing,uddin2015api}. Architecture consistency tools address architectural drift, checking whether implementation, deployment, or recovered architecture matches stated architectural descriptions~\cite{ali2018architecture,keim2023detecting}. Installation and dependency consistency checkers address stale environment assumptions, verifying that setup instructions match actual build configurations and dependency manifests~\cite{borovits2022findici,gao2025adapting}.
% GROUNDING: tan2024detecting -- "DOCER: regex identifier extraction + presence check across two repository snapshots; stale = present at first commit, absent at HEAD." (EMSE 29(1):5, 2024)
% GROUNDING: tan2007icomment,tan2012tcomment,panthaplackel2021deep -- "Comment/code consistency: iComment and @tComment detect comment-code inconsistencies; Panthaplackel et al. add deep just-in-time detection."
% GROUNDING: zhou2017analyzing,uddin2015api -- "API documentation consistency: directive defects between docs and code (Zhou et al.); empirical study of how API documentation fails (Uddin & Robillard)."
% GROUNDING: ali2018architecture,keim2023detecting -- "Architecture consistency: state-of-practice survey (Ali et al.); traceability-link recovery to detect architecture-doc inconsistencies (Keim et al.)."
% GROUNDING: borovits2022findici,gao2025adapting -- "Installation/dependency consistency: FindICI detects code/NL inconsistencies in IaC (Borovits et al.); installation-instruction adaptation (Gao et al.)."

AI configuration artifacts can exhibit all of these classes of divergence: they describe not only specific code elements but also architectural patterns, API tool parameters, dependency versions, and development conventions. The common thread across these consistency problems is that a claim made in a text artifact about the state of software can be operationalized as a query against the actual software, then flagged if it diverges.

%%% ============================================================
\section{Preliminary Study}
\label{sec:study}
%%% ============================================================

In this preliminary study, we focus on one specific form of context rot: referential rot, where a configuration file refers to functions, classes, constants, scripts, or file paths that no longer exist in the repository.

\subsection{Dataset}

We use the configuration files from the dataset of \citeauthor{galster2026dataset}~\cite{galster2026dataset}. After excluding empty files and pointer files (files that merely reference another file), and retaining only files with a recorded first commit SHA (required for the two-snapshot comparison), \CANDIDATEFILES{} files from \CANDIDATEREPOS{} repositories remain as candidates.
% GROUNDING: galster2026dataset -- "Input corpus of configuration files with first-commit SHAs; we filter to 8,213 files / 4,420 repos that have a recorded first commit." (arXiv 2605.08435)

We randomly sample \TOTREPOS{} repositories (random seed 42), retaining all configuration files present in each repository. This means that a repository containing both a \texttt{CLAUDE.md} and an \texttt{AGENTS.md} contributes both files to the analysis. The sample size was chosen to be statistically representative of the \CANDIDATEREPOS{} eligible repositories at a 95\% confidence level and a 5\% margin of error; we therefore interpret repository level estimates as estimates for the eligible dataset rather than only for the sampled repositories. The sample includes all major configuration file types present in the dataset.

\subsection{Applying DOCER to AI Configuration Files}

We apply \toolname{}'s detection logic as follows for each configuration file:

\begin{enumerate}
  \item \textbf{Clone.} We clone the repository locally.
  \item \textbf{Extract.} We run \toolname{}'s regular expressions against the current (HEAD) version of the configuration file using \texttt{git grep -howIP -f regex\_list.txt}, extracting candidate code elements.
  \item \textbf{Verify at first commit.} We search for each candidate in the source code at \texttt{first\_commit\_sha} (when the configuration file was first committed) using \texttt{git grep -howFI}, excluding README files and all AI configuration file types from the search scope.
  \item \textbf{Verify at HEAD.} We repeat the search at the current HEAD.
  \item \textbf{Classify.} Elements present at first commit but absent at HEAD are \emph{stale}. Elements present at HEAD are \emph{valid}. Elements absent from both snapshots are \emph{noise} and discarded.
\end{enumerate}

For this short paper, we deliberately do not tune \toolname{} for AI configuration files. Its regular expressions, verification strategy, and staleness definition are left unchanged. This allows us to ask what an existing documentation consistency checker can already reveal about AI configuration artifacts. Because we preserve \toolname{}'s traditional two-snapshot design, our measurement captures only references that were valid when the configuration file was first committed and later disappeared from the source. References introduced in subsequent edits to the configuration file fall outside this scope; we return to this and the opposing false-positive bias when interpreting the results.

\subsection{Results}

Table~\ref{tab:results} summarizes findings across \TOTREPOS{} repositories and \TOTCONFIGFILES{} configuration files. \toolname{}'s regular expressions extracted \CANDIDATEELEMS{} candidate code elements in total.

\begin{table}
  \caption{Stale code element references detected across \TOTREPOS{} randomly sampled repositories.}
  \label{tab:results}
  \begin{tabular}{lrr}
    \toprule
    & \textbf{Count} & \textbf{\%} \\
    \midrule
    Repositories analyzed & \TOTREPOS{} & N/A \\
    \quad with $\geq$1 stale reference & \STALEREPOS{} & \STALEREPOSPCT{} \\
    Configuration files analyzed & \TOTCONFIGFILES{} & N/A \\
    Candidate elements extracted & \CANDIDATEELEMS{} & N/A \\
    Verified references (found at first commit) & \VALIDELEMS{} & 100 \\
    \quad still valid at HEAD & \STILVALID{} & \STILVALIDPCT{} \\
    \quad stale (present then, absent now) & \STALEELEMS{} & \STALEPCT{} \\
    \bottomrule
  \end{tabular}
\end{table}

\toolname{} detects \STALEELEMS{} stale code element references across \STALEREPOS{} repositories, or \STALEREPOSPCT{}\% of those analyzed (95\% CI \STALEREPOSCILO{}--\STALEREPOSCIHI{}\%). Among the \STALEREPOS{} affected repositories, the median number of stale references was~1 and the maximum was~20. Table~\ref{tab:bytype} breaks them down by configuration file type. Stale references appear across all major types, with per-reference stale rates between 1.0\% and 1.4\%; these differences are small, and we do not test them for significance.

\begin{table}
  \caption{Verified and stale references by configuration file type. Copilot instructions includes \texttt{copilot-instructions.md} and other \texttt{*.instructions.md} files.}
  \label{tab:bytype}
  \begin{tabular}{lrrrr}
    \toprule
    \textbf{Config file type} & \textbf{Files} & \textbf{Verified} & \textbf{Stale} & \textbf{Stale \%} \\
    \midrule
    \texttt{CLAUDE.md}       & 147 & 5{,}423  & 77  & 1.42 \\
    \texttt{AGENTS.md}       & 234 & 6{,}762  & 70  & 1.04 \\
    Copilot instructions     & 211 & 5{,}436  & 77  & 1.42 \\
    \texttt{GEMINI.md}       &   6 &   133    &  1  & 0.75 \\
    \texttt{.cursorrules}    &   9 &   127    &  0  & 0.00 \\
    Other                    &   5 &   167    &  5  & 2.99 \\
    \midrule
    \textbf{Total}           & 612 & 18{,}048 & 230 & 1.27 \\
    \bottomrule
  \end{tabular}
\end{table}

To assess how well \toolname{} detects genuine context rot in this new setting, one author manually inspected a random sample of 50 elements classified as stale. As Table~\ref{tab:manual} shows, 32 (64\%) were genuine referential rot and the remainder false positives or ambiguous; five representative genuine cases appear in Table~\ref{tab:examples}. The false positives arise from \toolname{}'s broad regular expressions matching common English words or generic tokens that coincidentally appeared in the source at some point. They are informative for the transfer argument: they mark which parts of traditional documentation consistency tooling work directly here and which would benefit from artifact-specific tuning, motivating RQ2 (Section~\ref{sec:agenda}). As a threat to validity, the inspection was performed by a single annotator; inter-rater agreement was not assessed.

\begin{table}
  \caption{Manual inspection of 50 randomly sampled elements classified as stale.}
  \label{tab:manual}
  \begin{tabular}{lrr}
    \toprule
    \textbf{Classification} & \textbf{Count} & \textbf{\%} \\
    \midrule
    Genuine referential rot & 32 & 64 \\
    False positive                  & 12 & 24 \\
    Ambiguous                       &  6 & 12 \\
    \bottomrule
  \end{tabular}
\end{table}

More broadly, the \STALEREPOSPCT{}\% figure carries two opposing biases. \toolname{}'s two-snapshot criterion captures only references that were valid at the configuration file's first commit and misses any introduced by later edits, which deflates the estimate. Its broad regular expressions also match non-code tokens, so 36\% of elements classified as stale were false positives or ambiguous on inspection, which inflates it. We therefore read \STALEREPOSPCT{}\% as a feasibility signal rather than a precise prevalence; our sample also covers only public GitHub repositories that contain AI configuration files.

For practitioners, this check is available today~\cite{tan2023wait}. The lightweight two-snapshot \texttt{git grep} procedure used here needs no new tooling and runs in continuous integration, flagging configuration file references that have vanished from the source. The drift it surfaces is concrete, such as a renamed function, a deleted script, or a dropped dependency (Table~\ref{tab:examples}). Treating configuration files as code, and reviewing them alongside refactors that rename or delete elements, prevents much of this rot at the source.

\begin{table*}
  \caption{Five representative stale code element references detected in AI configuration files. Each element was present in the repository source at the configuration file's first commit but absent at HEAD.}
  \label{tab:examples}
  \begin{tabular}{p{3.2cm}p{2.9cm}p{3.5cm}p{1.4cm}p{3.2cm}}
    \toprule
    \textbf{Repository} & \textbf{Config file} & \textbf{Stale element} & \textbf{Type} & \textbf{Observed drift} \\
    \midrule
    \texttt{microsoft/pr-metrics} & \texttt{copilot-instruc\-tions.md} & \texttt{tsyringe} & library & Absent from source and build config at HEAD \\
    \addlinespace
    \texttt{dagu-org/dagu} & \texttt{ui/CLAUDE.md} & \texttt{runDAGs} & function & Function absent at HEAD (apparently renamed during UI rewrite) \\
    \addlinespace
    \texttt{smartystreets/}\newline\texttt{smartystreets-}\newline\texttt{python-sdk} & \texttt{CLAUDE.md} & \texttt{send\_risk\_lookup} & function & Method absent from SDK at HEAD \\
    \addlinespace
    \texttt{rolldown/rolldown} & \texttt{AGENTS.md} & \texttt{packages/rolldown/}\newline\texttt{src/binding.d.ts} & file path & File path absent at HEAD \\
    \addlinespace
    \texttt{EricLBuehler/}\newline\texttt{mistral.rs} & \texttt{AGENTS.md} & \texttt{scripts/}\newline\texttt{convert\_awq\_}\newline\texttt{marlin.py} & script & Script absent from repository at HEAD \\
    \bottomrule
  \end{tabular}
\end{table*}

%%% ============================================================
\section{Research Roadmap}
\label{sec:agenda}
%%% ============================================================

Our results show that referential rot appears in the sampled repositories and can be detected with existing tooling. The preliminary study addresses only one part of a larger problem. We identify four further transfer opportunities where prior work in documentation consistency maps onto open problems in AI configuration artifacts, summarized in Table~\ref{tab:roadmap}.

\begin{table*}
  \caption{Documentation consistency techniques and their analogues in AI configuration artifacts. The first row is the present study; the remaining rows are open transfer opportunities (RQ2). ``Existing tools'' names representative prior work; full citations appear in Section~\ref{sec:background} and below.}
  \label{tab:roadmap}
  \small
  \begin{tabular}{p{3cm}p{3.4cm}p{5.4cm}p{2.7cm}}
    \toprule
    \textbf{Consistency technique} & \textbf{Existing tools} & \textbf{AI configuration analogue} & \textbf{Transfer status} \\
    \midrule
    Referential (code/README) & \toolname{} & Code elements (functions, files, constants) that no longer exist & Direct; shown here \\
    \addlinespace
    Code comment & iComment, @tComment, deep JIT detection & Behavioral instructions that reference renamed or deleted elements & Open (RQ2) \\
    \addlinespace
    API documentation & Directive defect and API documentation checkers & MCP tool descriptions that diverge from their implementation & Open (RQ2) \\
    \addlinespace
    Architecture & Traceability link recovery & Architectural claims that drift as the system evolves & Open (RQ2) \\
    \addlinespace
    Installation / dependency & FindICI; installation instruction adaptation & Runtime and dependency versions that no longer hold & Open (RQ2) \\
    \bottomrule
  \end{tabular}
\end{table*}

Each opportunity pairs an established technique with an open problem (Table~\ref{tab:roadmap}): code comment consistency checkers~\cite{tan2007icomment,tan2012tcomment,panthaplackel2021deep}, API documentation checkers~\cite{zhou2017analyzing,uddin2015api}, architecture consistency tools~\cite{ali2018architecture,keim2023detecting}, and installation and dependency checkers~\cite{borovits2022findici,gao2025adapting}. This list is not exhaustive; AI configuration files also contain natural language descriptions of team conventions that are hard to operationalize as code queries. But it marks concrete starting points where prior work already exists.
% GROUNDING: tan2007icomment,tan2012tcomment,panthaplackel2021deep -- "Comment-code consistency checkers (grounded in Section 2.2); analogue = behavioral instructions referencing renamed/deleted elements."
% GROUNDING: zhou2017analyzing,uddin2015api -- "API documentation checkers (grounded in Section 2.2); analogue = MCP tool descriptions diverging from implementation."
% GROUNDING: ali2018architecture,keim2023detecting -- "Architecture consistency tools (grounded in Section 2.2); analogue = architectural claims that drift."
% GROUNDING: borovits2022findici,gao2025adapting -- "Installation/dependency checkers (grounded in Section 2.2); analogue = runtime/dependency versions that no longer hold."

We now state the research questions this agenda raises.

\paragraph{RQ1: What kinds of context rot occur in AI configuration artifacts?}
Context rot extends well beyond stale code element references. AI configuration files also contain architectural claims, tool use guidance, dependency and runtime assumptions, and behavioral conventions. Establishing the full taxonomy in practice requires large-scale, longitudinal analysis of the \citeauthor{galster2026dataset} corpus~\cite{galster2026dataset}, coding how configuration files diverge from their repositories over time. Our preliminary study addresses one subcategory, referential rot; the full taxonomy motivates the transfer opportunities in Table~\ref{tab:roadmap}.
% GROUNDING: galster2026dataset -- "Large-scale corpus of AI configuration artifacts across 4,738 repositories (9,470 configuration files among them) suitable for longitudinal divergence analysis." (arXiv 2605.08435)

\paragraph{RQ2: Which traditional documentation consistency techniques transfer to AI configuration artifacts without modification?}
\toolname{} transfers directly for referential rot; each remaining opportunity in Table~\ref{tab:roadmap} requires evaluation. Do code comment checkers flag stale behavioral instructions? Do API documentation checkers detect drift between MCP tool descriptions and their implementations? Each transfer attempt can follow the recipe used here: apply the unmodified tool to a corpus of the relevant configuration files, measure how often it fires, and manually validate precision. The result is an empirical estimate of transferability and evidence about which techniques work directly, which need tuning, and which require new approaches.

\paragraph{RQ3: Which forms of context rot affect AI assistant behavior?}
A configuration file can be stale without affecting the AI tool's output. Conversely, if the AI acts on a stale reference and generates code that imports a deleted module, the cost is real. Measuring this calls for controlled experiments comparing AI output quality across repositories with and without injected or naturally occurring context rot, extending the design of \citeauthor{lulla2026impact}~\cite{lulla2026impact}, who compare agents run with and without an \texttt{AGENTS.md}, to injected stale references. Such experiments may reveal that some categories (file paths, function names, library names) matter more than others.
% GROUNDING: lulla2026impact -- "Provides a with/without-AGENTS.md controlled design (runtime/token outcomes) extensible to injected stale references." (arXiv 2601.20404)

\paragraph{RQ4: How can detected context rot be repaired or mitigated?}
Deterministic detection and language model repair are complementary. A tool like \toolname{} identifies the specific stale element; a language model can then inspect the git history or the current codebase to suggest the updated reference. This hybrid sidesteps the precision concerns of LLM-only detection while keeping the flexibility needed for natural language repair, and it can be evaluated on whether the suggested replacement matches the element's actual rename in the history. Mitigation can also be preventive, alerting developers when a configuration file is not updated alongside related code changes.

%%% ============================================================
\section{Conclusion}
\label{sec:conclusion}
%%% ============================================================

AI configuration files guide coding assistants through the codebases, tools, architectures, and conventions of the projects they assist with. As software evolves, these descriptions go stale, a phenomenon we call context rot. Rather than building entirely new detection techniques, we argue that existing documentation consistency tools are a ready starting point: checkers for README files, code comments, API documentation, architecture descriptions, and installation instructions all address forms of divergence with direct analogues in AI configuration artifacts. As preliminary evidence, we applied \toolname{}, a consistency checker built for README files and wikis, to AI configuration files without modification, finding stale code element references in \STALEREPOSPCT{}\% of \TOTREPOS{} repositories. We identify four further transfer opportunities (Table~\ref{tab:roadmap}): behavioral instructions, MCP tool descriptions, architectural claims, and dependency references. As AI coding assistants become standard development infrastructure, keeping their configuration consistent with the evolving codebase is a software quality problem the community already has the tools to address.

\section*{Data Availability}
The online appendix~\cite{treude_2026_20588740} provides the analysis script, full per-element results, the repository versions analyzed, DOCER's regular expressions, the list of \STALEELEMS{} stale references, and the 50 manual annotations. \citeauthor{galster2026dataset} describe the input dataset of AI configuration files~\cite{galster2026dataset}; the dataset is archived on Zenodo~\cite{baltes_2026_20090356}.
% GROUNDING: galster2026dataset,baltes_2026_20090356 -- "Input dataset paper and Zenodo archive; Zenodo doi:10.5281/zenodo.19375880."

\balance

\end{document}